\documentstyle[11pt]{article}
\def\thefootnote{\fnsymbol{footnote}}
\def\be {\begin{eqnarray}}
\def\ee {\end{eqnarray}}
\def\beq {\begin{equation}}
\def\eeq {\end{equation}}
\def\slash{\hskip -.27cm /}

\def\np {Nucl. Phys.}
\def\prl {Phys. Rev. Lett.}
\def\pr {Phys. Rev.}
\def\pl {Phys. Lett.}
\def\tD {\tilde{D}}
\def\tD' {\tilde{D}^\prime}

\def\del {\partial}
\def\bi {\begin{itemize}}
\def\ei {\end{itemize}}
\def\ben {\begin{enumerate}}
\def\een {\end{enumerate}}

\def\Tr {{\rm Tr}}
\def\aa{{a_{I=1}^{KN}}}
\def\ab{{a_{I=0}^{KN}}}
\def\barB {\bar{B}}
\def\barN {\bar{N}}

\def\barK {\bar{K}}
\def\L{{\cal L}}
\def\lrarrow {\stackrel{\leftrightarrow}{{\partial}}_t}
\def\larrow{\stackrel{\leftarrow}{{\partial}}_t}
\def\rarrow{\stackrel{\rightarrow}{{\partial}}_t}
\begin{document}
\setcounter{page}{1}
\hfill {NORDITA-93/30 N}
\vskip -0.1cm
\hfill {SUNY-NTG-93-7}

\vskip 0.4in
\begin{center}
{\large\bf FROM KAON-NUCLEAR INTERACTIONS\\
TO KAON CONDENSATION}
\end{center}

\vskip 0.6in
\centerline{G.E. Brown$^{a}$\footnote{Supported in part by the Department
of Energy under Grant No. DE-FG02-88ER40388}, Chang-Hwan Lee$^{b}$\footnote
{Supported by the Korea Science and Engineering Foundation through the CTP
of Seoul National University},
Mannque Rho$^{c}$ and Vesteinn Thorsson$^{d}$}

\vskip 0.4in
\centerline{$^{a}$ {\it Department of Physics, State University of New York}}
\centerline{\it Stony Brook, N.Y. 11794, U.S.A.}

\centerline{$^{b}$ {\it Department of Physics, Seoul National University}}
\centerline{\it Seoul 151-742, Korea}

\centerline{$^{c}$ {\it Service de Physique Th\'{e}orique, C.E. Saclay}}
\centerline{\it F-91191 Gif-sur-Yvette, France}

\centerline{$^{d}$ {\it NORDITA, Blegdamsvej 17, DK-2100 Copenhagen \O,
Denmark}}

\vskip .6in

\centerline{\bf ABSTRACT}
\vskip .5cm
\noindent
An effective chiral Lagrangian in heavy-fermion formalism
whose parameters are constrained
by kaon-nucleon and kaon-nuclear interactions next to the leading order
in chiral expansion is used to describe kaon condensation in
dense ``neutron star" matter. The critical density is found to be
robust with respect
to the parameters of the chiral Lagrangian and comes out to
be $\rho_c\sim (3 - 4)\rho_0$. Once kaon condensation sets in, the system
is no longer composed of neutron matter but of nuclear matter.
Possible consequences on stellar collapse with the formation of compact
``nuclear stars" or light-mass black holes are pointed out.

\par\vfill

\newpage

\renewcommand{\thefootnote}{\arabic{footnote}}

\subsection*{Introduction}
\indent

Kaon condensation in dense nuclear and neutron matter predicted first
by Kaplan and Nelson\cite{kn} has recently been reformulated in chiral
perturbation theory (ChPT) to the leading chiral order by Politzer and
Wise \cite{pw} and by Brown et al \cite{bkrt} with results indicating
dramatic consequences in stellar collapse \cite{bkrt,vt,bb93}
and neutron star cooling
\cite{bkrt,vt}. In all these publications, the condensation process
is mainly driven by the attraction gained as the $KN$ sigma
term $\Sigma_{KN}$ associated with explicit breaking of chiral symmetry
\be
\Sigma_{KN}\approx \frac{1}{2}(\bar{m}+m_s)\langle N|(\bar{u}u+\bar{s}s)
|N \rangle \label{sigmaterm}
\ee
is ``rotated out" by nuclear density, where $\bar{m}=(m_u+m_d)/2$ is the mean
up-
and down-quark mass, $m_s$ the s-quark mass and $u, s$ are the relevant quark
fields. The $KN$ sigma term (\ref{sigmaterm}) which depends on
the s-quark content
of the proton could in principle be determined from experiment just as the
$\pi N$ sigma term could be determined from $\pi N$ scattering. Unfortunately
neither is the strangeness content of the proton
known nor are $KN$ scattering data accurate
enough to determine the quantity $\Sigma_{KN}$. If we assume
as an extreme case that $\langle \bar{s}s\rangle_N\equiv
\langle N|\bar{s}s|N \rangle\approx 0$, then
we obtain $\Sigma_{KN}\sim 1.3 m_\pi$ at the tree order with the chiral
Lagrangian we write down below. If on the other hand we take
$\langle \bar{s}s\rangle_N/\langle (\bar{u}u +\bar{d}d)\rangle_N
\approx 0.2$ which some of current hadrons models predict, we get
$\Sigma_{KN}\sim 3.7 m_\pi$. In the literature, this range of magnitude
of the sigma term has led to what appears to be a fairly robust prediction
for kaon condensation at a matter density three or four times nuclear matter
density $\rho_0$ ($\approx \frac 12 m_\pi^3$). We shall consider this
to be the range of values to keep in mind in what follows, although generally
favoring the value of
\be
\Sigma_{KN} \sim 6\Sigma_{\pi N}\sim 2m_\pi\label{eq2}
\ee
using the empirical value of the $\pi N$ sigma term. The choice of this value
will be justified later.

The problem with this prediction based on the scalar attraction due to
(\ref{sigmaterm}) is that by itself such an attraction is not compatible
with available data on kaon-nucleon and kaon-nuclear interactions at low
energy. As pointed out by several authors \cite{critique}, there seems to
be no serious need for a remnant of the scalar attraction in S-wave $K^{\pm}N$
and $K^{\pm}$-nuclear scattering amplitudes with the bulk of the data
adequately accounted for by standard $\rho$- and $\omega$-meson-exchange
mechanisms \cite{speth}.

The purpose of this paper is to show that kaon condensation
{\it does still occur} at the same low density, ($\sim (3 - 4) \rho_0$), with
an
effective chiral Lagrangian that is compatible with low-energy
kaon-nuclear interactions. The critical density could be lowered
even further if one invokes the previously proposed in-medium
scaling (which we shall call ``BR scaling") in the chiral Lagrangians
\cite{br91}. We shall do this
using heavy-fermion chiral perturbation expansion \cite{JM}
to next to the leading order. We shall consider a chiral Lagrangian
consisting of the octet baryons $B=B^a T^a$
and the octet pseudoscalar mesons $\pi=\pi^a T^a$ in heavy-fermion formalism
(HFF) \cite{JM}, first in free space and then in medium. We will focus on
S-wave kaon-baryon interactions, so the decuplet baryons which we expect
to be unimportant for this partial wave will not be
included in our consideration. Although vector mesons do not figure explicitly
in our effective Lagrangian, their chiral symmetry and scaling properties
in nuclear medium will be invoked as in ref.\cite{br91} to infer the scaling
behavior of the relevant term in the Lagrangian.

\subsection*{Chiral Perturbation Theory}
\indent

In arranging terms in a {\it consistent} chiral expansion for mesons and
baryons, it proves to be convenient to use the Weinberg counting
rule \cite{wein91} according to which an amplitude involving $E_N$ number
of external nucleon lines and $E_K$ number of external kaon lines can be
characterized by $Q^\nu$ in the amplitude
where $Q$ is the characteristic small momentum
scale involved in the process and
\be
\nu=2+2L-\frac 12 E_N +\sum_i \left(d_i +\frac 12 n_i -2\right)
\label{count}
\ee
where $L$ is the number of loops, the sum over $i$ goes over {\it all}
vertices, $d_i$ the number of derivatives that act
on the $i$th vertex and $n_i$ the number of nucleon lines attached to
the $i$th vertex. In the absence of external fields, chiral symmetry
constrains\footnote{In the presence of slowly varying electroweak
fields, the constraint is $P_i\geq -1$. This is important in nuclear
exchange currents \cite{pmr92}.}
\be
P_i\equiv d_i+\frac 12 n_i -2\geq 0.
\ee
Applied to $KN$ scattering, we see that the leading term in this counting
is given by $L=0$ and $P_i=0$ which is satisfied by a vertex with
$d_i=1$ and $n_i=2$, so the amplitude has the index $\nu=1$. At the next
order, we can have $L=0$ and one $P_i=1$ vertex with an index $\nu=2$.
{\it We note that no loops contribute to this order}. One-loop terms contribute
at the next order, say, $\nu=3$ together with tree graphs involving
one vertex with $d_i=2$. In this paper, we will limit ourselves
up to $\nu=2$ for kaon-nucleon scattering and hence no
loops need be calculated. Applied to kaon-nuclear scattering and
to kaon condensation, this procedure effectively
takes into account density-dependent
one-loop contributions consistently with chiral symmetry as discussed in
\cite{pw}.
This chiral counting is embodied in the chiral Lagrangian first
written down by Jenkins and Manohar \cite{JM} which we shall use:
\be
\L = \L_0 +\L^\prime,\label{lag}
\ee
where
\be
\L_0 &=& \frac{f^2}{4}\Tr \del_\mu U\del^\mu U^\dagger + r\Tr M \left(U
+h.c.-2 \right) \nonumber\\
&+& \Tr \barB iv\cdot D B
+2D\Tr \barB S^\mu \{A_\mu, B\}+2F\Tr \barB [A_\mu, B],\label{lag1}\\
\L^\prime &=&a_1\Tr \barB\left(\xi M \xi+ h.c.\right) B +a_2\Tr \barB B
\left(\xi M\xi
+ h.c.\right) +a_3\Tr \barB B \Tr\left(MU + h.c.\right)\nonumber\\
&+& c_1 \Tr \barB D^2 B +c_2\Tr \barB (v\cdot D)^2 B \nonumber\\
&+& d_1\Tr \barB A^2 B +d_2 \Tr \barB (v\cdot A)^2 B +d_3\Tr \barB B A^2
+d_4\Tr \barB B (v\cdot A)^2 \nonumber\\
&+&d_5 \Tr \barB B  \Tr A^2
+d_6 \Tr \barB B \Tr (v\cdot A)^2
+d_7 \Tr \barB A_\mu \Tr A^\mu B \nonumber\\
&+&d_8 \Tr \barB (v\cdot A) \Tr (v\cdot A) B
+d_9 \Tr \barB A_\mu B A^\mu
+d_{10} \Tr \barB (v\cdot A) B (v\cdot A)
\nonumber\\
&+& f_1 \Tr \barB (v\cdot D)(S\cdot A)B +f_2\Tr \barB (S\cdot D)(v\cdot A)B
+f_3\Tr \barB[S^\alpha,S^\beta]A_\alpha A_\beta B\nonumber\\
&+& \cdots\label{lag2}
\ee
where $v^\mu$ is the four-velocity of the heavy baryon (with $v^2=1$) and
\be
D^\mu B &=& \del^\mu B +[V^\mu,B],\\
V^\mu &=& \frac 12 (\xi\del^\mu \xi^\dagger +\xi^\dagger\del^\mu \xi), \ \ \
A^\mu=\frac{1}{2i}(\xi\del^\mu\xi^\dagger -\xi^\dagger\del^\mu\xi)
\ee
with $\xi^2=U$. Here $S^\mu$ is the spin operator $S^\mu=\frac 14\gamma_5
[\not\!{v},\gamma^\mu]$ constrained to $v\cdot S=0$, $a_i$, $c_i$, $d_i$ and
$f_i$ are parameters to be fixed later. Except for some minor differences,
we use the notations of Jenkins and Manohar\cite{JM} where the advantage of
using this Lagrangian for chiral perturbation theory is clearly pointed out.
This Lagrangian was recently shown to be suited for describing
chiral properties of
nuclear systems by Park et al\cite{pmr92} where one can also find detailed
discussions on how to systematically compute  higher chiral-order
terms using this Lagrangian.

We now illustrate the counting rule discussed above in terms of the
Lagrangian (\ref{lag}). The leading order $KN$ (say $K^+ N$ )
scattering is described
by $\L_0$. Specifically the $\nu=1$ S-wave $KN$ amplitude is given by
$i\Tr\barB[V^0,B]$ which can be written explicitly as
\be
\L_{\nu=1}=\frac{-i}{8f^2}\left(3(\barN \gamma^0 N)\barK\lrarrow K
+(\barN\vec{\tau}\gamma^0 N)\cdot\barK\vec{\tau}\lrarrow K\right)\label{lnu1}
\ee
with $N^T=(p\ \  n)$, $K^T=(K^+\  K^0)$ and $\barK\lrarrow K\equiv
\barK \rarrow K-\barK\larrow K$. For
$\barK N$ scattering, due to G-parity, the isoscalar term changes sign.
In terms of an effective Lagrangian
that contains vector mesons such as hidden gauge symmetry Lagrangian
of ref.\cite{bando},
the first term of (\ref{lnu1}) can be identified as the $\omega$ exchange
and the second term as the $\rho$ exchange between the kaon and the nucleon.
Thus we can think of the leading-order contribution as vector-dominated.
This is in agreement with the standard phenomenological meson-exchange
picture \cite{speth}. Now
the next chiral order amplitude comes from $\L^\prime$ (\ref{lag2}).
For S-wave $KN$ scattering, the terms with the coefficients $c_i$ and
$f_i$ do not contribute, so the relevant part of the Lagrangian simplifies
(for S-wave) to
\be
\L_{\nu=2} &=& \frac{\Sigma_{KN}}{f^2} (\barN N) \barK K
+\frac{C}{f^2} (\bar{N}\vec{\tau} N)\cdot (\bar{K}\vec{\tau}K)\nonumber\\
& & +\frac{\tilde{D}}{f^2}
(\barN N)\del_t \barK\del_t K+ \frac{\tilde{D}^\prime}{f^2} (\barN \vec{\tau}
N)\cdot(\del_t \barK\vec{\tau}\del_t K)\label{lnu2}
\ee
where
\be
\Sigma_{KN}&= &-(\frac 12 a_1+a_2 +2a_3) m_s,\\
C &=& -\frac{a_1 m_s}{2},\\
\tilde{D} &=& \frac 14(d_1+d_2+d_7 +d_8)+\frac 12(d_3 +d_4)+(d_5+d_6),\\
\tilde{D}^\prime &=& \frac 14 (d_1+d_2+d_7+d_8). \label{dprime}
\ee
As stated, the Lagrangian (\ref{lnu1},\ref{lnu2})
at tree order gives the leading amplitude with $\nu=1$ and the next-to-leading
amplitude with $\nu=2$.
Loops with (\ref{lag1}) give $\nu\geq 3$ and loops involving (\ref{lag2})
even higher. Thus follows the statement that
to order $\nu=2$ there are no loop contributions.
This makes the calculation simpler than in $\pi\pi$ or $KK$ interactions
where the next-to-leading order corrections involve both higher derivative
counter terms and one-loop terms. This counting result in HFF is manifestly
simpler than the relativistic formulation of ChPT \cite{gss}
where numerous counter terms --
none of which can be determined from other processes -- intervene.

In the strategy of chiral perturbation theory,
the constants that appear in (\ref{lnu1}) and (\ref{lnu2}) are
to be determined from experiments. As mentioned above,
the $KN$ sigma term $\Sigma_{KN}$ could, in principle, be determined
from low-energy $KN$ scattering data or from the expression (\ref{sigmaterm})
once we know the strangeness content of the nucleon $\langle N|\bar{s}s|N
\rangle$. At the moment the sigma term is not known.  We will take the value
quoted above for our discussion. As for the coefficients $\tilde{D}$ and
$\tilde{D}^\prime$ we anticipate an important contribution from
the $1/m_B$ correction due to baryon-antibaryon pair terms.
These ``$1/m_B$" corrections  can be readily
evaluated in the formalism (see Park {\it et al}\ \cite{pmr92}).
The argument is sketched in the Appendix.  Here we simply
give the results
\be
\tilde{D}_{\frac{1}{m}} &\approx& -
\frac{1}{48} \left[(D+3F)^2+9(D-F)^2\right]/m_B
\approx -0.12/m_B\approx -0.024\,{\rm fm},\nonumber\\
\tilde{D}^\prime_{\frac{1}{m}} &\approx& -\frac{1}{48}\left[(D+3F)^2
-3(D-F)^2\right]\approx -0.086/m_B\approx -0.017\,{\rm fm}
\label{paircon}
\ee
where $m_B$ is the (centroid) baryon mass which we take $\sim 1$ GeV.
We have used here the tree-level fits $F=0.44$ and
$D=0.81$. There will also be contributions that arise from degrees of
freedom whose mass scale is higher than the chiral expansion scale
$\Lambda_\chi\sim 1$ GeV. These can appear as a counter-term contribution
and cannot be calculated in the scheme. They should be determined from
experiments. In some cases such as in $\pi \pi$ scattering, those constants
can be saturated by resonances (for instance, the constants $L_i$
in the Lagrangian of $O(Q^4)$ in the pion sector are dominated
by the vector mesons $\rho$, $a_1$ etc.) but here while this is plausible,
we have not succeeded in finding such a simple mechanism.

\subsection*{On-Shell Constraints}
\indent

What can we say about the constants $\tilde{D}$ and $\tilde{D}^\prime$
from experiments?

For the purpose of determining these constants from experiments, it
is simpler to look at $K^+ N$ scattering. The $K^- N$ scattering is
somewhat more delicate because of the resonance $\Lambda (1405)$.
It has been suggested \cite{critique,speth,buttgen} that the presently
available
data on $K^{\pm}$-nucleon and $K^{\pm}$-nuclear scattering
indicate that the bulk of the data can be understood reasonably well by
the leading $\nu=1$ term (\ref{lnu1}) {\it with the $\nu=2$ terms effectively
suppressed}. Let us see what this means in the framework of chiral
perturbation theory. At present the data on threshold $K^+ N$ scattering,
particularly the isoscalar amplitudes, are not
good enough to allow a precise determination of the constants.
We can nonetheless make a simple analysis which is still meaningful
as described below.

 From the effective Lagrangian (\ref{lnu1}) and (\ref{lnu2}), we
can immediately write down the expressions for
scattering lengths
\be
a_{I=1}^{KN} &=&\frac{1}{4\pi f^2(1+m_K/m_B)}\left(-m_K+\Sigma_{KN}+
C+(\tilde{D}+\tilde{D}^\prime)m_K^2\right),\label{a1}\\
a_{I=0}^{KN}&=& \frac{1}{4\pi f^2(1+m_K/m_B)}\left(\Sigma_{KN}-3C+
(\tilde{D}- 3\tilde{D}^\prime)m_K^2 \right).\label{a0}
\ee
where $a^{KN}_{I=0,1}$ are the S-wave scattering
lengths for $K^+ N$ scattering in isospin $I=0,1$.
Given experimental values for $a_{I=0,1}^{KN}$ and knowing $\Sigma_{KN}$,
these equations could determine the constants $\tilde{D}$ and
$\tilde{D}^\prime$.

Scattering lengths have recently been obtained from a rather complete analysis
by Barnes and Swanson \cite{barnes}. They find
\be
a_{I=1}^{KN} &\approx& -0.31 {\mbox{fm}},\nonumber\\
a_{I=0}^{KN} &\approx& -0.09\  {\mbox{fm}}.\label{emp}
\ee
Whereas $\aa$ is reliably determined, $\ab$ is highly uncertain: It
can differ between 0
and $-0.2$ fm \cite{VPI}\footnote{T. Barnes, private communication.}.

As long as $\ab$ is small in magnitude compared with $\aa$, the nuclear
scattering of kaons should be insensitive to $\ab$. This suggests
a procedure which minimizes the dependence on the precise value of
$\ab$. The isospin averaged amplitude
\be
\bar{a}=\frac 34 \aa + \frac 14 \ab
\ee
will occur in kaonic interactions in nuclear matter. In the case of
neutron star matter, another combination occurs, but we will show that
this latter case is insensitive to how we handle the on-shell kaon scattering.
{}From (\ref{emp}), we find the empirical value
\be
\bar{a}_{emp}\approx -0.255\, {\rm fm},\label{empnuc}
\ee
which is rather insensitive to the precise value of $\ab$ within the given
limits.

M\"{u}ller-Groeling {\it et al} \cite{buttgen} have analyzed
$\bar{K}N$ scattering in
the framework of boson exchange. They employ vector mesons with roughly
the same $SU(3)$ coupling coefficients as are implicitly incorporated in
eq.(\ref{lnu1}). Whereas the scalar attraction between kaon and nucleon is
built into the chiral Lagrangian (\ref{lnu2}) through the $\Sigma_{KN}$
term, M\"{u}ller-Groeling {\it et al} \cite{buttgen} include an exchange of an
explicit scalar particle with
\be
g_{\sigma\mbox{\tiny  NN}} g_{\sigma \mbox{\tiny KK}}/4\pi\approx 0.9
\ee
and $m_\sigma\approx 600\, {\rm MeV}$. Using eqs. (30) and (31)
of Brown, Koch and Rho \cite{bkr}, one can convert our sigma-term
attraction into an effective $\sigma$-exchange by
\be
\left(\frac{g_{\sigma\mbox{\tiny NN}} g_{\sigma\mbox{\tiny KK}}}{4\pi}
\right)_{eff}
=\frac{1}{4\pi} \frac{m_\sigma^2\Sigma_{KN}}{2f^2 m_K}\approx 0.93
\ \label{attraction}
\ee
where we have used $\Sigma_{KN}\approx 2m_\pi$ and $f=f_\pi \approx 93$ MeV
in the conversion.
The agreement between the M\"{u}ller-Groeling {\it et al} value and ours is
quite satisfactory. We thus see that there is evidence for attractive scalar
interaction in the $\bar{K}N$ scattering consistent with $\Sigma_{KN}\approx
2m_\pi$. \footnote{The hypothetical higher-mass scalar repulsion
introduced by B\"{u}ttgen {\it et al}\ \cite{buttgen} in the $K^+ N$ channel
is probably simulating the so-called ``counter-term" contribution to
$\tilde{D}$ and $\tilde{D}^\prime$ in the chiral Lagrangian discussed below
and may be understood in the chiral bag language as a ``van der Waals'
repulsion" discussed by Vento {\it et al}\ \cite{vento}.}

To see the role of the constants $\tilde{D}$ and $\tilde{D}^\prime$,
let us first set them
equal to zero. Then from (\ref{a1}) and (\ref{a0}), we would predict
(for $\Sigma_{KN}\approx 2 m_\pi$)
\be
a_{I=1}^{KN} &\approx& -0.30\, {\rm fm},\label{aa1}\\
a_{I=0}^{KN} &=& \frac{\Sigma_{KN}-3C}{4\pi f^2 (1+m_K/m_B)}
\approx 0.46\, {\rm fm}.\label{aa0}
\ee
We have made clear in (\ref{aa0}) that in the absence of the
$\tilde{D}$ and $\tilde{D}^\prime$ corrections, the $a_{I=0}^{KN}
$ is given {\it mainly} by the explicit chiral symmetry breaking.
For this amplitude, the vector meson amplitude is zero. Equations
(\ref{aa1}) and (\ref{aa0}) lead to $\bar{a}_{th}\approx -0.11$ fm which
is too small compared
with the empirical value (\ref{empnuc}). Including the baryon-antibaryon
pair contribution (\ref{paircon}) brings $a_{I=0}^{KN}$ down to
$0.40\ {\rm fm}$, increasing $\bar{a}_{th}$  to $\bar{a}_{th}
\approx -0.15 {\rm fm}$. Now if in addition to the pair contributions,
one decreases the value of $\Sigma_{KN}$
by 30\%, then one gets $\bar{a}\approx -0.255\, {\rm fm}$. Therefore it is
not difficult to fit the amplitude $\bar{a}$ within the uncertainty of the
parameters of the chiral Lagrangian. However it will then be
difficult to understand in this way the scattering lengths
$\ab$ and $\aa$ {\it separately}, even within the wide range of
allowed values in $\ab$.

Suppose we ignore the $\tilde{D}$ and $\tilde{D}^\prime$ terms but account
for higher-order effects in the constants of the chiral Lagrangian.
The most prominent quantity is $f$. Now at one loop ({\it i.e.},
at $\nu=3$), the kaon field will scale with the kaon decay constant
$f_K$, so $f$ should be replaced
by $f_K$ in the sigma term $\sim \Sigma_{KN}/f^2$.
On the other hand, the
$\nu=1$ term (\ref{lnu1}) involves vector-meson exchange and requires
that $f$ be identified with $f_\pi$, the pion decay constant.
Calculation of other $\nu=3$ contributions has not yet been carried out,
so it is difficult to quantify our argument
but the use of $f_K$ instead of $f_\pi$ in the explicit symmetry breaking
-- which is equivalent to using $(\Sigma_{KN})_{eff}\approx 0.69 \Sigma_{KN}$
-- generally moves our theoretical values towards the central values of the
empirical ones. In fact a somewhat lower value
\be
(\Sigma_{KN})_{eff}\approx 0.6 \Sigma_{KN}
\ee
would lead to an agreement between $\bar{a}_{th}$ and $\bar{a}_{emp}$.

Let us now see what values of $\tilde{D}$ and $\tilde{D}^\prime$
are required to reproduce the ``empirical" results (\ref{emp}). {}From
(\ref{a1}) and (\ref{a0}), we find
\be
\tilde{D}&\approx& 0.33/m_K-\Sigma_{KN}/m_K^2,\label{D}\\
\tilde{D}^\prime &\approx& 0.16/m_K-C/m_K^2 .\label{D'}
\ee
For $\Sigma_{KN}\approx 2 m_\pi$, we have $\tilde{D}\approx -0.47/m_B$
with $m_B\approx 2 m_K$. The
constant $\tilde{D}^\prime$ is $ \approx 0.46/m_B$ independently of the value
of the sigma term. (If one takes $\ab=0$, one gets $\tilde{D}\approx
-0.41/m_B$ and $\tilde{D}^\prime\approx 0.38/m_B$ while for $\ab= -0.2\
{\rm fm}$, one finds $\tilde{D}\approx -0.57/m_B$ and $\tilde{D}^\prime
\approx 0.54/m_B$.)
Comparing with (\ref{paircon}), one sees that the pair term
accounts for only a small part of the constants $\tilde{D}$ and
$\tilde{D}^\prime$, the latter disagreeing even in sign.
The conclusion then is that there must be additional corrections
to the scattering lengths at order $\nu\geq 2$.
These corrections can be of two different classes: One class would
be of the $\nu=2$ terms coming from higher energy sector that is integrated
out and is not saturated by the exchange of single resonances and
the other would be loop corrections. As for the first class, we have no clue
as to its mechanism. All we can do is to extract it from
experiments. As for the second, since,
as mentioned, there are no loop corrections to the $\tilde{D}$ and
$\tilde{D}^\prime$ terms, they must enter with
$\nu\geq 3$, contributing effectively to the scattering lengths as do
$\tilde{D}$ and $\tilde{D}^\prime$ terms. Such calculations are in progress
and will be reported elsewhere.

That loops could be important either in $\pi N$ scattering \footnote{This
possibility was discussed in a somewhat different language by Delorme
{\it et al} \ \cite{delorme} and also by K. Kubodera and H. Yabu (private
communication).} or in $KN$ scattering or in both is pretty much obvious.
In $\pi N$ scattering, our chiral Lagrangian (\ref{lag})
gives, at the $\nu=2$ order, the isoscalar $\pi N$ scattering length
$$ a^+_{\pi N}\equiv \frac 12 \left(a^{\pi^+ p} +a^{\pi^+ n}\right)
=[4\pi f^2 (1+m_\pi/m_B)]^{-1} \left(2\tilde{D}_{\pi\mbox{\tiny N}}
m_\pi^2 +\Sigma_{\pi N}\right)$$
where $\Sigma_{\pi N}$ is the $\pi N$ sigma term $\approx 45$ MeV and
$$\tilde{D}_{\pi\mbox{\tiny N}}=\frac 14 (d_1+d_2) +\frac 12 (d_5+d_6).$$
Now if one takes the empirical value of $a^+_{\pi N}$ \cite{delorme},
$a^+_{\pi N}=-0.01 m_\pi^{-1}$, then one must have
$$\tilde{D}_{\pi\mbox{\tiny N}}\approx -0.27\, {\rm fm} \,.$$
It should be noted, however, that the empirical value of
$a^+_{\pi N}$ can be reproduced by just taking the pair contribution
{\it alone} (which is just a part of the $\tilde{D}$ term in the Lagrangian),
or, stated differently, by the pseudovector pole contribution\cite{delorme},
or equivalently by the contribution of nucleon recoil\cite{schwinger}.
This means in our scheme that
loop contributions must be present such as to substantially cancel
the contribution of $\Sigma_{\pi N}$ to the scattering length\cite{delorme}.
This could be explained by a mechanism that enhances the pair contribution
to $\pi N$ scattering by order $\sim 2 m_N/m_\pi$ \footnote{
We conjecture that this enhancement may be understood by a mechanism
analogous to the Nambu-Jona-Lasinio picture of the pion in terms of
$N\bar{N}$ bubbles. In effective field theories such as ours,
such a phenomenon could
occur -- as in BCS theory -- through a collective mechanism which turns
``irrelevant" terms (of higher chiral order) into
``marginal" terms. See Polchinski \cite{pol} on a discussion of this matter
in the framework of effective field theories.}.

In what follows, we will simply use the {\it empirical} values (\ref{D})
and (\ref{D'}) effectively {\it parametrizing} $\nu\geq 2$ effects
quadratic in kaon frequency. Later we will find that $K^-$ condensation
is little affected by these $\nu\geq 2$ effects. The reason can be seen from
(\ref{lnu2}). Since the $\tilde{D}$ and $\tilde{D}^\prime$ multiply
$\del_t K \del_t \bar{K}$, these corrections scale with density as
$\omega_{\mbox{\tiny K}}^2/m_{\mbox{\tiny K}}^2$, where
$\omega_{\mbox{\tiny K}}$ is the kaon frequency. The loop corrections
will also scale in the same way or what is more likely, even faster.
The $\omega_{\mbox{\tiny K}}$ decreases
with increasing density, both because the kaon experiences the attractive
scalar field of (\ref{attraction}) and because of the attractive vector field
from $\omega$-meson exchange with the nucleon described by the first term
of (\ref{lnu1}). Thus the terms prefixed by $\tilde{D}$ and $\tilde{D}^\prime$
on the right-hand side of eq.(\ref{lnu2}) (and loop corrections absorbed in
them) decrease compared with the first term with the coefficient
$\Sigma_{KN}/f^2$. Consequently although the influence of the $\Sigma_{KN}/f^2$
term on the scattering lengths may be cut by $\sim 40$ \% by the corrections,
once the densities necessary for kaon condensation are reached, the influence
of the correction terms will be greatly diminished.

In relativistic heavy-ion collisions, in which there is not enough time for
strangeness violation, $\bar{K}K$ must be created \cite{HI}. Interactions of
kaons and nucleons via $\omega$-meson exchange have opposite signs for $K$
and $\bar{K}$, so that they average out. Interactions via $\rho$-meson
exchange go out for isospin-symmetric matter. The baryon-antibaryon pair
term and loop corrections do not decrease with increasing density compared
with $\Sigma_{KN}/f^2$ term, so that the corrections may become important.
This issue would require a more careful treatment of higher chiral corrections.

\subsection*{In-Medium Scaling}
\indent

So far we have been considering S-wave kaon interactions in free space.
We would now like to take the chiral Lagrangian so defined and apply
it to kaon interactions in {\it nuclear matter}. For this we have to account
for the effect of the medium on chiral symmetry and other symmetries of
QCD incorporated into the effective chiral Lagrangian.
As argued by Brown and Rho
\cite{br91}, the most economical way to implement chiral symmetry and
trace anomaly of QCD -- which are the most important properties of
QCD at low energy -- in nuclear medium is to endow to the
in-medium effective chiral Lagrangian with the basic parameters of the theory
that {\it scale} as a function
of density. Arguments based on symmetries of QCD predict a universal
scaling (valid at the mean field with the Lagrangian) of the
{\it quasiparticles} relevant for the process in question \cite{br91}:
\be
\frac{m_B^*}{m_B}\approx \frac{m_M^*}{m_M}\approx \frac{f^*}{f}
\label{brscaling}
\ee
where the subscripts $B$ and $M$ stand, respectively, for baryons and mesons
in the $SU(2)$ sector
(except for Goldstone bosons) and the star denotes in-medium quantities at
a density $\rho\neq 0$. The unstarred quantities denote free-space
quantities. At the mean-field level, the constant $g_A$ remains
unscaled \cite{br91}. There is by now rather strong evidence that the scaling
(\ref{brscaling}) is valid in nuclei \cite{expbr}.
The consequence of this in-medium scaling on the effective Lagrangian
(\ref{lnu1}) and (\ref{lnu2}) is simply that we replace $f$ by $f^*$.
In medium, one is effectively including one-loop terms as mentioned above.

As alluded above, to one-loop order in free space,
the kaon decay constant $f_K$ receives $O(Q^2)$ corrections relative
to the pion decay constant $f_\pi$.
So working to that level in medium, $f_K$ should be distinguished
from $f_\pi$ in the effective Lagrangian, not only in its magnitude but also
in its behavior. This could be implemented with a Lagrangian consisting
of Goldstone bosons and baryons only. One possible way is to write,
following Gasser and Leutwyler \cite{GL}, a symmetry breaking term
that involves derivatives of $O(M\partial^2)$, where $M$ is the quark mass
matrix, and then perform a one-loop ChPT with this Lagrangian. This is
very similar to what one does in skyrmion physics except that here only tree
terms are considered. In this paper we shall not pursue this procedure.
We find it far more transparent when the light
vector mesons $\rho$, $\omega$, etc are explicitly present.
Now in medium, we have $f_\pi\rightarrow f_\pi^*\approx \Phi (\rho) f_\pi$
whereas model calculations (such as NJL) indicate that $f_K$ scales
very little up to $\rho\approx \rho_0$, so we could assume $f^*_K\approx
f_K$ up to nuclear matter.\footnote{While there are no reliable lattice
calculations in dense matter, we know what happens to the strange quark
condensate $\langle \bar{s}s\rangle_0$ in high temperature. The calculation
by Kogut {\it et al}\ \cite{kogut} shows that the strange-quark condensate
changes little as a function of temperature while the light-quark condensate
does. This is consistent with our assumption.}
This may be somewhat too naive but it should
be more reasonable than taking it to scale like $f_\pi^*$.
The consequence of this argument is then that in medium $f$ in eq.(\ref{lnu1})
is replaced by $f_\pi^*$ and that in eq.(\ref{lnu2}) by $f_K$.
The replacement in (\ref{lnu1}) can be understood best in
the description with explicit vector mesons. As noted in \cite{br91},
the gauge coupling of the vector meson $g$ remains unscaled at the mean-field
level, so the scaling of $f_\pi^*$ means the scaling of the vector meson mass
$m_V^*$ by the KSRF relation $m_V^*=2{f_\pi^*}^2 g^2$.
Thus the factor $1/{f^*}^2$ just corresponds to $2g^2/{m_V^*}^2$ for
the vector propagator in the medium at zero momentum transfer. We
will simply assume that neither $\Sigma_{KN}$ nor $\tilde{D}$ nor
$\tilde{D}^\prime$
scales with density. In reality, there may be some scaling in both: the former
because it is related to the condensate difference between the vacuum
and the hadronic ``bag"; the latter because at least part of it may
come from the pair term which depends inversely on the centroid
baryon mass $m_B$ which presumably scales.
We expect however that this is a fine-tuning that cannot be done accurately,
so we will not pursue it any further.

The remaining procedure for describing kaon condensation in neutron star
matter is identical to what was done in \cite{bkrt,vt}, so we will be very
brief. Details are found in refs.\cite{bkrt,vt,tpl}. We simply mention that
in medium, the Lagrangians (\ref{lnu1}) and (\ref{lnu2}) generate one-loop
terms and that kaon condensation will be triggered by the presence of
electrons.

\subsection*{Equation of State with Kaon Condensation}
\indent

The equation of state (EOS) describing the state of matter containing kaon
condensates can be evaluated at the mean field level using the Lagrangian
(\ref{lag}) with (\ref{lag1}) and (\ref{lag2}), retaining all nonlinear
meson interactions. For this, loop corrections may be quantitatively important.
In contrast, evaluating the critical density is relatively simple as it is
insensitive to nonlinearities and loop corrections. Here we give the results
for critical density and composition of the condensed matter computed with
a chiral Lagrangian consistent with the on-shell constraints to the chiral
order $\nu=2$.

Before implementing the $\tilde{D}$ and $\tilde{D}^\prime$ corrections,
let us recall the previous results of refs.\cite{bkrt,vt,tpl}
obtained in the leading chiral order, $\nu=1$ supplemented by a $KN$ sigma
term. The critical density is given as the nucleon density at which
the pole of the kaon propagator $D$ is equal to to the electron chemical
potential in the absence of the condensate. The electron chemical potential
is determined by the nuclear matter equation of state and the conditions of
beta equilibrium and local charge neutrality. Equivalently,the critical density
is the nucleon density at which the energy density is lowered by the
introduction of kaon condensate. In Table 1
\footnote{With $a_1 m_s=-67$ MeV and $a_2 m_s=125$ MeV determined by
Gell-Mann-Okubo mass formulas, the parameter
$a_3 m_s=-134$, $-222$, $-310$ MeV corresponds
respectively to $\Sigma_{KN}= 1.3 m_\pi$, $2.5 m_\pi$, $3.8 m_\pi$. The
case $a_3 m_s=-134$ MeV corresponds to $\langle \bar{s}s\rangle_N/
\langle \bar{u}u\rangle_N \approx 0.07$. This represents the least favorable
condition for kaon condensation and will be used for illustrating the equation
of state we predict. Note that
the difference from eq.(\ref{eq2}) is related to the well-known $\Sigma_{\pi
N}$
problem in the tree order of chiral Lagrangians, which is resolved by
going to higher orders in chiral perturbation theory. In the present
problem, it is best to consider the numerical value of ``$\Sigma_{KN}$" as
a parameter.}, we quote the results for the
critical density without BR scaling as given in refs.\cite{vt,tpl}, which are
essentially those of \cite{bkrt} but with the inclusion of muons in beta
equilibrium with the neutron star matter. Here as well as in what follows,
we use the potential contribution to the symmetry energy $F(u)=u$
\cite{bkrt,tpl,pal}.
\begin{center}
\bigskip
{\bf TABLE 1}\\
\begin{quotation}
{\noindent The critical densities
without $\tilde{D}$ and $\tilde{D}^\prime$ terms and without BR scaling.
The parameters held fixed are: $a_1 m_s=-67$ MeV, $a_2 m_s= 125$ MeV,
$f_K\approx f_\pi\approx 93$ MeV.}
\end{quotation}
\bigskip
\begin{tabular}{lrrr} \hline \hline
$F(u)$     &   $a_3m_s$(MeV)     &     $\,\,\,\,u_c\,\,\,$  \\ \hline
           &   - 134            &     4.11\\
$u$        &   - 222             &     3.04\\
           &   - 310             &     2.39\\ \hline\hline
\end{tabular}
\end{center}
\vskip 0.25 cm

Next we incorporate $\tilde{D}$ and $\tilde{D}^\prime$ terms so as to
be consistent (at the order $\nu=2$) with the $K^+ N$ scattering lengths.
In what follows, we will simply take (\ref{D}) and (\ref{D'}) valid to
order $\nu=2$. A more sophisticated analysis would require more accurate
$K^+ N$ data and loop corrections. The inverse propagator is
\be
D^{-1}(\omega) &=& \omega^2 - m_K^2 - \Pi(\omega) \nonumber\\
&=& ( 1 +[\frac{\tilde{D}}{f^2} + (2x-1) \frac{\tilde{D}^\prime}
{f^2}]u \rho_0 )\omega^2 - m_K^2
+ u \rho_0\frac{1+x}{2f^2} \omega
+ \frac{u \rho_0}{f^2}(\Sigma^{K N}+(2x-1)C).
\nonumber\\
\label{dm1}
\ee
\bigskip

\begin{center}
\bigskip
{\bf TABLE 2}\\
\begin{quotation}
{\noindent The critical densities in ChPT with $\nu=2$ terms
without ($u_c^{ns}$) and with ($u_c^{s1,s2}$) BR scaling.
For parameter values, see Table 1.}
\end{quotation}
\bigskip
\begin{tabular}{lrrrrr} \hline \hline
$F(u)$     &   $a_3m_s$(MeV)    & $u_c^{ns}$& $u_c^{s1}$  & $u_c^{s2}$\\ \hline
           &   - 134            &   4.20   & 2.84 & 3.15\\
$u$        &   - 222            &   3.27 & 2.49 & 2.69\\
           &   - 310            &   2.60 & 2.16 & 2.39\\ \hline\hline
\end{tabular}
\end{center}
\bigskip

\noindent
The critical density computed with eq.(\ref{dm1}) {\it without BR scaling}
is given in Table 1, denoted as $u_c^{ns}$. To implement the BR scaling
in eq.(\ref{dm1}), we replace $f_\pi\rightarrow f_\pi^*$ and
$f_K\rightarrow f_K\approx f_\pi$ in the manner prescribed above.
(We are disregarding the $O(Q^2)$ correction to $f_K$ in our numerical
estimates.) As argued above, this corresponds to scaling
the vector interaction only. Now if we take the scaling to be
$f_\pi^* = f_\pi / ( 1 + 0.25u ) $,  we get the critical density labeled
$u_c^{s1}$ in Table 2.
The results are only slightly changed if one uses a slower
scaling, say, $f_\pi^* = f_\pi / ( 1 + 0.16u ) $. They are
given in Table 2 as $u_c^{s2}$.

The composition of the condensed matter is illustrated in Tables 3 and 4.
In Table 3 are given the results obtained {\it without BR scaling} and
in Table 4 those {\it with BR scaling} with
$f_\pi^*/f_\pi\approx (1+0.25 u)^{-1}$.
We have shown only the results with the small $\Sigma_{KN}\approx 1.3 m_\pi$
(or $a_3 m_s=-134$ MeV) to illustrate the robustness of the prediction.
The quantities reported in the Tables are:
The chiral angle $\theta$, the energy density gain $\Delta\epsilon$
(in MeV/fm$^3$)
, the chemical potential $\mu$ (in MeV), the proton fraction
$x$, the kaon fraction $x_K=\rho_K/\rho$, the electron fraction
$x_e=\rho_e/\rho$ and the muon fraction $x_\mu=\rho_\mu/\rho$ as a
function of matter density $u=\rho/\rho_0$.

\vskip 0.5cm
\begin{center}
\bigskip
{\bf TABLE 3} \\
\begin{quotation}
{\noindent
``Nuclear star" composition without BR scaling for $\Sigma_{KN}\approx
1.3 m_\pi$ ($a_3 m_s\approx -134$ MeV).}
\end{quotation}
\bigskip
\begin{tabular}{rrrrrrrr} \hline \hline
     u  &$\theta$&$\Delta \epsilon$&$\mu$&$x$&$x_K$&$x_e$&$x_\mu$ \\
\hline
   4.20 &   0.4 &    0.0 &   256.7 &  0.195 &  0.000 &  0.111 &  0.085 \\
   4.70 &  31.6 &   -3.0 &   223.9 &  0.291 &  0.180 &  0.066 &  0.046 \\
   5.20 &  43.6 &  -11.2 &   191.3 &  0.360 &  0.301 &  0.037 &  0.022 \\
   5.70 &  51.7 &  -23.7 &   159.9 &  0.409 &  0.380 &  0.020 &  0.009 \\
   6.20 &  57.4 &  -39.3 &   130.8 &  0.444 &  0.431 &  0.010 &  0.002 \\
   6.70 &  61.5 &  -57.4 &   104.5 &  0.468 &  0.464 &  0.005 &  0.000 \\
   7.20 &  64.5 &  -77.1 &    81.1 &  0.486 &  0.484 &  0.002 &  0.000 \\
   7.70 &  66.8 &  -98.3 &    60.4 &  0.499 &  0.499 &  0.001 &  0.000 \\
   8.20 &  68.6 & -120.5 &    42.0 &  0.509 &  0.509 &  0.000 &  0.000 \\
   8.70 &  70.0 & -143.5 &    25.8 &  0.517 &  0.517 &  0.000 &  0.000 \\
   9.20 &  71.1 & -167.3 &    11.3 &  0.522 &  0.522 &  0.000 &  0.000 \\
   9.70 &  72.0 & -191.6 &    -1.7 &  0.526 &  0.526 &  0.000 &  0.000 \\
  10.20 &  72.7 & -216.5 &   -13.4 &  0.530 &  0.530 &  0.000 &  0.000 \\
 \hline\hline
\end{tabular}
\end{center}
\bigskip

Some remarkable features in the results are as follows.
\begin{itemize}
\item The critical densities are robust with respect to the
parameters of the effective Lagrangian. The on-shell constraints
bring only small modification. The in-medium scaling reduces the critical
density from $u_c\sim 4$ to $u_c\sim 3$.
\item With the in-medium scaling, the critical density is fairly
insensitive to the numerical value of $\Sigma_{KN}$.
\item Once kaon condensate sets in, the initially dense neutron matter
turns quickly to nuclear matter, with the proton fraction $x$ being
already quite substantial just above the critical density. This shows that with
kaon condensates, a compact star is more likely a ``nuclear star"
rather than neutron star of the standard scenario.
\item The kaon fraction becomes equal to the proton fraction slightly above
the critical density; the proton charge is almost entirely balanced by
the $K^{-}$ charge.
\end{itemize}

\begin{center}
\bigskip
{\bf TABLE 4}\\
\begin{quotation}
{\noindent ``Nuclear star" composition with BR scaling with $f_\pi^*/f_\pi
=(1+0.25 u)^{-1}$ and $\Sigma_{KN}\approx 1.3 m_\pi$ ($a_3 m_s\approx
-134$ MeV.)}
\end{quotation}
\bigskip
\begin{tabular}{rrrrrrrr} \hline \hline
     u  &$\theta$&$\Delta \epsilon$&$\mu$&$x$&$x_K$&$x_e$&$x_\mu$ \\
\hline
   2.84 &   0.0 &    0.0 &   209.6 &  0.147 &  0.000 &  0.089 &  0.058 \\
   3.34 &  28.0 &   -7.0 &   134.5 &  0.356 &  0.331 &  0.020 &  0.005 \\
   3.84 &  30.2 &  -19.5 &    92.8 &  0.424 &  0.419 &  0.006 &  0.000 \\
   4.34 &  29.9 &  -33.3 &    66.0 &  0.457 &  0.455 &  0.002 &  0.000 \\
   4.84 &  29.0 &  -47.5 &    47.6 &  0.475 &  0.474 &  0.001 &  0.000 \\
   5.34 &  27.8 &  -61.7 &    34.6 &  0.485 &  0.485 &  0.000 &  0.000 \\
   5.84 &  26.6 &  -76.1 &    25.3 &  0.491 &  0.491 &  0.000 &  0.000 \\
   6.34 &  25.5 &  -90.5 &    18.4 &  0.495 &  0.495 &  0.000 &  0.000 \\
   6.84 &  24.4 & -105.0 &    13.2 &  0.498 &  0.498 &  0.000 &  0.000 \\
   7.34 &  23.3 & -119.6 &     9.3 &  0.499 &  0.499 &  0.000 &  0.000 \\
   7.84 &  22.4 & -134.4 &     6.3 &  0.501 &  0.501 &  0.000 &  0.000 \\
   8.34 &  21.5 & -149.3 &     4.0 &  0.501 &  0.501 &  0.000 &  0.000 \\
   8.84 &  20.6 & -164.5 &     2.2 &  0.502 &  0.502 &  0.000 &  0.000 \\
   9.34 &  19.9 & -179.9 &     0.8 &  0.502 &  0.502 &  0.000 &  0.000 \\
   9.84 &  19.2 & -195.4 &    -0.2 &  0.502 &  0.502 &  0.000 &  0.000 \\
 \hline \hline
\end{tabular}
\end{center}

\subsection*{Conclusion: Astrophysical Consequences}
\indent

Our results show that the critical density is remarkably
robust. It comes out to be $u_c\sim (3 - 4) $ quite independently of the
constraints from kaon-nuclear interactions. What is also noteworthy
is that once the BR scaling is implemented, the dependence
on the strangeness content of the proton is considerably weaker.
Even when one assumes that $\langle N|\bar{s}s|N\rangle =0$ which is
certainly unreasonably conservative, the critical density is
$u_c <3$. This robust nature of the condensation leads to
some striking consequences on stellar collapse.

Kaon condensation plays an important role in the collapse of large stars,
once central densities exceed the critical density $u_c$. Brown {\it et al}
\cite{bkrt} showed that electrons changed into $K^-$-mesons and neutrinos in
a fraction of a second, the neutrinos being trapped for a longer time (of
$\sim$10 seconds). Since a large fraction of protons are present in  kaon
condensation, nuclear matter rather than neutron matter -- formed in the
conventional scenario -- is reached and this has the effect of substantially
softening the  dense matter equation of state (EOS) because of lower
symmetry energy of nuclear matter. Because of the greater binding energy,
more energy will be emitted in neutrinos \cite{tpl}.

Effects of kaon condensation on the supernova explosion mechanism remain
to be explored. However effects of kaon condensation on the structure of
neutron stars and on some aspects of the explosion have been worked out
by Thorsson {\it et al}\,\cite{tpl}.

Because kaon condensation softens the nuclear EOS at high densities,
it substantially diminishes the maximum mass $M_{max}$ for neutron
stars. Brown \cite{geb92} found, with reasonable assumption about
the compression modulus of nuclear matter that
\be
M_{max}\approx 1.5 M_{\odot}.
\ee
This has major implications for the formation of black holes in
stellar collapse \cite{bb93}. Stars in the range of $\sim 18$ to
30 $M_{\odot}$ can first explode, returning matter to the galaxy
and then go into low-mass black holes of mass $M_{\mbox{\tiny BH}}
\sim 1.5 M_{\odot}$. Upon collapse, stars heavier than $\sim 30
M_{\odot}$ drop directly into black holes without nucleosynthesis.
These black holes are heavier, of mass $M_{\mbox{\tiny BH}}\geq
10 M_{\odot}$. Brown and Bethe \cite{bb93} estimate that
$\sim 10^9$ black holes have been formed in the above way in the
galaxy.

\subsection*{Acknowledgments}

We are grateful for useful correspondence from K. Kubodera, M. Prakash
and H. Yabu
on related problems and for helpful comments from W. Weise.
One of us (CHL) would like to acknowledge valuable
discussions with H. Jung with whom further work is in progress.

\newpage
\subsection*{Appendix}
\renewcommand\theequation{A.\arabic{equation}}
\setcounter{equation}{0}
Here we sketch briefly the derivation of the $1/m_B$ corrections. Consider the
lowest
order Lagrangian with baryon fields,
\be
{\cal L}=\Tr\bar B\left(i D \slash  -m(1- v\;\slash)B+
{\rm F} \gamma^\mu\gamma_5 [A_\mu, B]
+{\rm D} \gamma^\mu\gamma_5 \{A_\mu, B\} \right) \;\; .
\ee
Rewriting the fields as
$  B = \frac{1}{\sqrt{2}} B^a \lambda^a $, $ V_\mu = V_\mu^a \lambda^a $ and
$  A_\mu = A_\mu^a \lambda^a $, where $\lambda^a ,\, a=1,8$ are the
Gell-Mann matrices,
the Lagrangian can be written
\be
{\cal L} = \bar B^a i D \slash^{\,\,a c} B^c -m \bar B^a \delta^{ac}
(1-v\;\slash)B^c + \bar B^a \gamma^\mu\gamma_5 {\cal A}^{ac}_\mu B^c .
\ee
Here
\be
i D \slash^{\,\,ac}&=& i\delta^{ac}\partial\; \slash +i (2i f^{abc}) V \slash^b
   \nonumber \\
{\cal A}^{ac}_\mu &=& {\rm F} (2i f^{abc}) A^b_\mu +{\rm D}(2 d^{abc}) A^b_\mu
\ee
and $f^{abc}, d^{abc}$ are $SU(3)$ structure constants.
The equation of motion for the heavy baryon field is
\be
\left(g^{ac} -m ( 1-v\;\slash) \delta^{ac}\right) B^c =0
\label{eq}
\ee
with $ g^{ac} = i D \slash^{\,\,ac} +\gamma^\mu\gamma_5 {\cal A}^{ac}_\mu $.
Now, decompose $B^a$ using
$ P_+ =\frac{1}{2}(1+v\;\slash)$ and $ P_- =\frac{1}{2}(1-v\;\slash)$,
\be B^a=P_+ B^a + P_- B^a =B^a_+ + B^a_-\; .
\ee
Applying $P_-$ to eq.(\ref{eq}) from the left, we have
\be B^c_-=\frac{1}{2m}P_- g^{ce} B^e_+ +O(\frac{1}{m^2})\;\; .
\label{bm}\ee
Now applying  $P_+$ to eq.(\ref{eq}) and using eq.(\ref{bm}),
we get, modulo $O(1/m^2)$,
\be
P_+\left( g^{ae}+\frac{1}{2m}g^{ac} P_- g^{ce} \right) B^e_+=0 \;\; .
\ee
Thus the Lagrangian containing  $1/m_B$ correction term is
\be
{\cal L} =\bar B^a_+ ( g^{ae}+\frac{1}{2m} g^{ac} P_- g^{ce}) B^e_+.
\ee
To write this explicitly, we note that the $O(A^2)$ terms are given by
\be
g^{ac} P_- g^{ce} =\gamma^\mu\gamma_5 {\cal A}^{ac}_\mu P_-\gamma^\nu\gamma_5
{\cal A}^{ce}_\nu = - {\cal A} \slash^{ac} P_+ {\cal A} \slash^{ce}
  = -(v\cdot {\cal A}^{ac}) (v\cdot {\cal A}^{ce}) \;\; ,
\ee
\noindent which leads to $(m=m_B)$
\be
{\cal L}_{1/m} \longrightarrow -\frac{1}{2m_B} \bar B^a (v\cdot {\cal A}^{ac})
  (v\cdot {\cal A}^{ce})B^e \;\; .
\ee
     {}From the identities
\be
(2i f^{abc})(2i f^{cde}) &=& \frac{1}{2} {\rm Tr} \left( \lambda^a [\lambda^b,
    [\lambda^d, \lambda^e]]\right) \;\; , \nonumber \\
(2i f^{abc})(2 d^{cde}) &=& \frac{1}{2} {\rm Tr} \left( \lambda^a [\lambda^b,
    \{\lambda^d, \lambda^e\}]\right) \;\; , \nonumber \\
(2 d^{abc})(2i f^{cde}) &=& \frac{1}{2} {\rm Tr} \left( \lambda^a \{\lambda^b,
    [\lambda^d, \lambda^e]\}\right) \;\; , \nonumber \\
(2 d^{abc})(2 d^{cde}) &=& \frac{1}{2} {\rm Tr} \left( \lambda^a \{\lambda^b,
    \{\lambda^d, \lambda^e\}\}\right) -\frac{2}{3} {\rm Tr}(\lambda^a\lambda^b)
   {\rm Tr}(\lambda^d\lambda^e) \;\; ,
\ee
we have
\be
\bar B^a (v\cdot {\cal A}^{ac})(v\cdot {\cal A}^{ce}) B^e
&=& \Tr\bar B\left( {\rm D}^2\{ v\cdot A,\{v\cdot A, B\}\}
   + {\rm D}{\rm F}\{v\cdot A,[v\cdot A, B]\}\right. \nonumber\\
 &&\left. +{\rm F}^2[v\cdot A,[v\cdot A,B]]+{\rm F}{\rm D}
[v\cdot A,\{v\cdot A,B\}] \right) \nonumber\\
 && -\frac{4}{3} {\rm D}^2 {\rm Tr}( \bar B v\cdot A) {\rm Tr}(v\cdot A
B)\nonumber\\
&=& ({\rm D}+{\rm F})^2\Tr\bar B(v\cdot A)^2 B
+ 2 ({\rm D}^2-{\rm F}^2)\Tr\bar B(v\cdot A)B(v\cdot A) \nonumber\\
&& +({\rm D}-{\rm F})^2\Tr\bar B B (v\cdot A)^2
  -\frac{4}{3} {\rm D}^2 {\rm Tr}( \bar B v\cdot A) {\rm Tr}(v\cdot A B) \; .
\nonumber\\ \,
\ee
For S-wave scattering,
\be
{\rm Tr} \bar B A_0^2 B &\longrightarrow &\frac{1}{4f^2} \left(
(\bar N N)(\partial_t \bar K
 \partial_t K)+ (\bar N \vec\tau N)\cdot (\partial_t\bar K \vec\tau
\partial_t K) \right)
\nonumber\\
{\rm Tr} (\bar B A_0){\rm Tr} (A_0 B) &\longrightarrow &\frac{1}{4f^2} \left(
(\bar N N)(\partial_t \bar K
 \partial_t K)+ (\bar N \vec\tau N)\cdot (\partial_t\bar K \vec\tau
\partial_t K) \right)
\nonumber\\
{\rm Tr} \bar B B A_0^2 &\longrightarrow & \frac{1}{2 f^2}
  (\bar N N)( \partial_t\bar K\partial_t K) \nonumber\\
{\rm Tr}\bar B A_0 B A_0 &\longrightarrow& 0 \;\; .
\ee
The $1/m_B$ corrections are therefore
\be
{\cal L}_{\frac{1}{m}}=\tilde{D} \frac{1}{f^2} \left(
\bar N N \partial_t \bar K \partial_t K\right)
+\tilde{D}^\prime \frac{1}{f^2}
  (\bar N \vec\tau N)\cdot (\partial_t\bar K \vec\tau \partial_t K) \;\; ,
\ee
with
\begin{eqnarray}
\tilde{D}_{\frac{1}{m}} &=& -\frac{1}{48 m_B}(({\rm D}+3{\rm F})^2 +
9({\rm D}-{\rm F})^2) 
 \label{d} \\
\tilde{D}^\prime_{\frac{1}{m}} &=& -\frac{1}{48 m_B} (({\rm D}+3 {\rm F})^2
-3 ({\rm D}-{\rm F})^2 ) \;\; .
 \label{e}
\end{eqnarray}


\newpage
\parindent 0 pt
\begin{thebibliography}{99}
\bibitem{kn}
D.B. Kaplan and A.E. Nelson, \pl \ {\bf B175} (1986) 57
\bibitem{pw}
H.D. Politzer and M.B. Wise, \pl \ {\bf B273} (1991) 156
\bibitem{bkrt}
G.E. Brown, K. Kubodera, M. Rho and V. Thorsson, \pl \ {\bf B291} (1992) 355
\bibitem{vt}
V. Thorsson, Ph.D. thesis, SUNY at Stony Brook (1992), unpublished
\bibitem{bb93}
G.E. Brown and H.A. Bethe, ``A scenario for a large number of low mass black
holes in the galaxy," CALTECH preprint,
submitted to Astrophysical Journal (1993)
\bibitem{critique}
See, {\it e.g.}, W. Weise, Proceedings Int. Nucl. Phys. Conf., Wiesbaden,
27 July-1 August 1992; \np \ {\bf A}, to be published
\bibitem{speth}
P.B. Siegel and W. Weise, \pr \ {\bf C38} (1988) 221
\bibitem{buttgen}
A. M\"{u}ller-Groeling, K. Holinde and J. Speth, \np \ {\bf A513} (1990) 557;
R. B\"{u}ttgen, K. Holinde, A. M\"{u}ller-Groeling, J. Speth and
P. Wyborny, \np \ {\bf A506} (1990) 586
\bibitem{br91}
G.E. Brown and M. Rho, \prl \ {\bf 66}, 2720 (1991)
\bibitem{JM}
E. Jenkins and A. Manohar, \pl \ {\bf B255} (1991) 558; {\bf B259} (1991)
353; E. Jenkins, \np \ {\bf B368} (1991) 190
\bibitem{wein91}
S. Weinberg, \pl \ {\bf B251} (1990) 288; \np \ {\bf B363} (1991) 2
\bibitem{pmr92}
T.-S. Park, D.-P. Min and M. Rho, ``Chiral dynamics and heavy-fermion
formalism in nuclei," submitted to Phys. Repts. (1993)
\bibitem{bando}
M. Bando, T. Kugo and K. Yamawaki, Phys. Repts. {\bf 164} (1988) 217
\bibitem{gss}
J. Gasser, M.E. Sainio and A. \u{S}varc, \np \ {\bf B307} (1988) 779
\bibitem{barnes}
T. Barnes and E.S. Swanson, MIT-CTP-2169/ORNL-CCIP-92-15
(December 1992)
\bibitem{VPI}
J.S. Hyslop, R.A. Arndt, L.D. Roper and R.L. Workman, \pr \ {\bf D46}
(1992) 961
\bibitem{bkr}
G.E. Brown, V. Koch and M. Rho, \np \ {\bf A535} (1991) 701
\bibitem{vento}
V. Vento, M. Rho and G.E. Brown, \pl \ {\bf B103} (1981) 285
\bibitem{delorme}
J. Delorme, M. Ericson and T.E.O. Ericson, \pl \ {\bf B291} (1992) 379
\bibitem{schwinger}
J. Schwinger, `` Particles and Sources '', Gordon and Breach, New York (1969)
\bibitem{pol}
J. Polchinski, ``Effective field theory and the Fermi surface,"
1992 TASI Lecture
\bibitem{HI}
A.E. Nelson and D.B. Kaplan, \pl \ {\bf B192} (1987) 193; C.M. Ko,
Z.G. Wu, L.H. Xia and G.E. Brown, \prl \ {\bf 66} (1991) 2577; G.E.
Brown, C.M. Ko, Z.G. Wu and L.H. Xia, \pr \ {\bf C43} (1991) 1881
\bibitem{kogut}
J.B. Kogut, D.K. Sinclair and K.C. Wang, \pl \ {\bf 263} (1991) 101
\bibitem{expbr}
C. Adami and G.E. Brown, Phys. Repts., to appear
\bibitem{GL}
J. Gasser and H. Leutwyler, Ann. Phys. (N.Y.) {\bf 158}, (1984) 142;
\np \ {\bf B307} (1988) 763;(1991) 353
\bibitem{tpl}
V. Thorsson, M. Prakash and J.M. Lattimer,
`` Composition, structure and evolution of neutron stars with kaon
condensates", NORDITA-93/29 N / SUNY-NTG-92-33
\bibitem{pal} M.Prakash, T. L. Ainsworth and J.M. Lattimer, \prl
\ {\bf 61} (1988) 2518
\bibitem{geb92}
G.E. Brown, Proc. First Symp. on Nucl. Phys. of the Universe, Oak Ridge,
TN, September 24-26, 1992, to be published.
\end{thebibliography}
\end{document}